%% file: main.tex
\documentclass[conference]{IEEEtran}
\IEEEoverridecommandlockouts

\input{Packages/packages}
\input{Intro/acronyms}
\begin{document}

	\input{Intro/title}

	\input{Intro/authors}

	\maketitle

	\input{Intro/abstract}

	\input{Intro/keywords}

	\bstctlcite{bistcontrol}

	\input{Sections/introduction}
	\input{Sections/experimentalSetUp}
	\input{Sections/Results/results}
	\input{Sections/conclusions}

	\bibliographystyle{IEEEtran}
	\bibliography{Bibliography/bibliography}

\end{document}

%% file: Packages/packages.tex
\usepackage{cite}
\usepackage{amsmath,amssymb,amsfonts}
\usepackage{algorithmic}
\usepackage{textcomp}
\usepackage{xcolor}
\usepackage{hyperref}
\usepackage[a4paper, total={184mm,239mm}]{geometry}
\usepackage{subfigure}
\usepackage[printonlyused,withpage]{acronym}

\usepackage{graphicx}

\usepackage{caption}

\def\BibTeX{{\rm B\kern-.05em{\sc i\kern-.025em b}\kern-.08em
    T\kern-.1667em\lower.7ex\hbox{E}\kern-.125emX}}

\renewcommand{\IEEEbibitemsep}{0pt plus 0.5pt}
\makeatletter
\IEEEtriggercmd{\reset@font\normalfont\fontsize{7.9pt}{8.40pt}\selectfont}
\makeatother
\IEEEtriggeratref{1}

%% file: Intro/acronyms.tex
\acrodef{SACRES}[SACRES]{Safety Critical Real-Time Embedded System}
\acrodef{DMR}[DMR]{Double Modular Redundancy}
\acrodef{AI}[AI]{Artificial Intelligence}
\acrodef{PMU}[PMU]{Performance Monitoring Unit}
\acrodef{HPC}[HPC]{Hardware Performance Counter}
\acrodef{ISA}[ISA]{Instruction Set Architecture}
\acrodef{SBU}[SBU]{Single-Bit Upsets}
\acrodef{MBU}[MBU]{Multiple-Bit Upsets}
\acrodef{intRF}[intRF]{Integer Register File}
\acrodef{SDC}[SDC]{Silent Data Corruption}
\acrodef{SMOTE}[SMOTE]{Synthetic Minority Oversampling Technique}
\acrodef{PCA}[PCA]{Principal Component Analysis}
\acrodef{HTDR}[HTDR]{Hard-To-Detect Region}
\acrodef{FC-FFNN}[FC-FFNN]{Fully Connected Feed Forward Neural Network}
\acrodef{LSTM}[LSTM]{Long-Short Term Memory}

%% file: Intro/title.tex
\title{
	Micro-Architectural features as soft-error induced fault executions markers in embedded safety-critical systems: a preliminary study
}

%% file: Intro/authors.tex
\author{
	\IEEEauthorblockN{Deniz Kasap\textsuperscript{+}, Alessio Carpegna\textsuperscript{*}, Alessandro Savino\textsuperscript{*}, Stefano Di Carlo\textsuperscript{*}}\\
	\IEEEauthorblockA{
		\textit{\textsuperscript{*}Politecnico di Torino, Control and Computer Engineering Department, Torino, Italy } \\
		\textit{\textsuperscript{+}Bilkent University, Ankara, Turkey} \\
		{Corresponding email: stefano.dicarlo@polito.it}
	}

%
%
%
%
%
}

%% file: Intro/abstract.tex
\begin{abstract}

Radiation-induced soft errors are one of the most challenging issues
in \ac{SACRES} reliability, usually handled using different flavors of \ac{DMR} techniques. 
This solution is becoming unaffordable due to the complexity of modern microprocessors in all domains.
This paper addresses the promising field of using \ac{AI} based hardware detectors for soft errors.
To create such cores and make them general enough to work with different software applications, microarchitectural attributes are a fascinating option as candidate fault detection features. Several processors already track these features through dedicated \ac{PMU}. However, there is an open question to understand to what extent they are enough to detect faulty executions.
Exploiting the capability of \textit{gem5} to simulate real computing systems, perform fault injection experiments and profile microarchitectural attributes (i.e., \textit{gem5} Stats), this paper presents the results of a comprehensive analysis regarding the potential attributes to detect soft error and the associated models that can be trained with these features. 
\end{abstract}

%% file: Intro/keywords.tex

%% file: Sections/introduction.tex
\section{Introduction}
\label{sec:indroduction}

Radiation-induced soft errors, which started as a rather exotic failure mechanism causing anomalies in satellites, have become one of the most challenging issues in all electronic systems, particularly \acf{SACRES} \cite{Heijmen:2010aa}. Many efforts have been spent in the last decades to measure \cite{8791555}, model \cite{7094277}, and mitigate \cite{Sayil:2019va} radiation effects, implementing cross-layer reliability countermeasures \cite{Vallero:2019aa}. 
Predictability is a crucial SACRES requirement as it helps ensure the system's safety and reliability and makes it easier to test and maintain. Predictability is implemented by static partitioning and hardware isolation of available computing resources (e.g., memory, CPU cores, etc.) among a set of predefined tasks, making it easier to monitor the behavior of each task in isolation. Resilience to soft errors is then supported through redundancy at different levels \cite{10.1145/3203217.3203240}. In particular, \acf{DMR} implementing lock-step execution is a popular schema to achieve fault detection, and check-pointing is the solution to enable recovery from faults \cite{8974333}. However, with the increasing complexity of microprocessor cores, \ac{DMR} is becoming unaffordable, and designers are increasingly looking into smaller hardware/software error detectors \cite{4711620,4408257,4484788,9962356}. \ac{AI} is a fascinating instrument in this domain, bringing to the new concept of artificial resilience, i.e., systems that can be trained to detect and possibly recover from faults \cite{10.1145/3535044.3535062}. 

\ac{AI} has been employed for constructing hardware or software soft error detectors trained on software-specific input/output features \cite{Khosrowjerdi:2018aa,Vishnu:2016aa,vassiliadis2021artificial,wang2018neural}. Overall, in these approaches, the data used to determine whether the output of a task is correct or incorrect (i.e., the feature vector) includes the task input and output. While effective, having an application-dependent feature vector makes it challenging to create generic detectors, especially when looking at hardware implementations that require standard methods to collect and deliver features to the detector. Microarchitectural features (i.e., executed instructions, cache misses, or incorrectly anticipated branching for the active program) are easier to collect thanks to the availability of increasingly complex \acf{PMU} tracking and measuring numerous performance-related events accessible through dedicated \ac{HPC}. Therefore an important question to investigate is: ``are microarchitectural features able to explain faulty executions in the presence of soft errors?". Dutto et al. showed that the answer to this question is 'yes' in the case of permanent faults \cite{Dutto2021}. However, permanent faults lead to fault accumulation during software execution, amplifying anomalies and making detection easier. \cite{da-Rosa:2019aa} confirms a positive answer to this question, even if a deep analysis of several features is not reported. Recently Nosrati et al. proposed an \ac{AI} detector for soft errors in embedded processors \cite{Nosrati:2022vn}. While results are promising in terms of fault detection accuracy, the approach relies on monitoring internal signals of the microprocessor that would require an invasive hardware redesign.

This paper proposes a preliminary study to understand to what extent microarchitectural features traced through a \ac{PMU}
can be exploited to build an \ac{AI}-powered hardware soft error detector. For this purpose, a set of fault injection experiments performed using \textit{FIMSIM} \cite{Yalcin:2011aa}, a fault injector framework based on \textit{gem5} \cite{Binkert:2011aa} was used to create a dataset of features over several faulty and correct executions. This dataset was analyzed to give designers insights into the best options to build their soft error detection systems. Particular emphasis was devoted to understanding whether event timing could bring additional information to the model. This information is crucial to identify the best \ac{AI} models to employ in this challenging task. The paper does not focus on the design of the detector but on assessing the feasibility of the idea. Therefore, the hardware implementation cost and related inference time are out of the scope of this study.

%% file: Sections/experimentalSetUp.tex
\section{Experimental set-up}
\label{sec:experimental-setup}

\begin{figure}[!ht]
			\centering
			\includegraphics[width=\columnwidth]
			{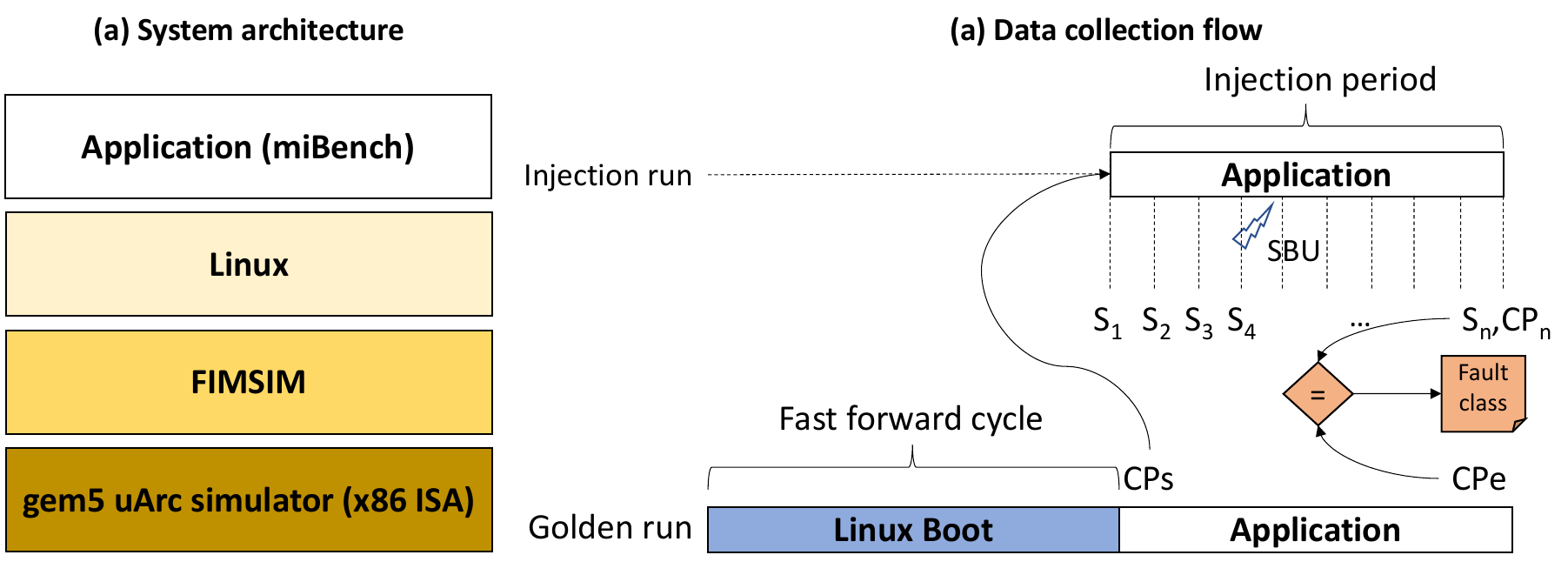}
			\caption{\small{Experimental setup based \textit{gem5} and \textit{FIMSIM}.}}
			\label{fig:setup}
		\end{figure}

Injecting faults directly into real hardware is complex \cite{6962073,9962356}; virtual simulators simplify the development of fault injection modules and the subsequent data collection task \cite{Papadimitriou:2021aa,Chatzidimitriou:2016aa,Benso:2007aa}. For this reason, simulation-based fault injection was used to collect data. \autoref{fig:setup} summarizes the basic building blocks of the implemented experimental design.  
		
Since the goal is to analyze microarchitecture level features, the \textit{gem5} simulator was used to emulate the hardware substrate \cite{Binkert:2011aa}. 
Experiments were performed by simulating the full-system stack, modeling the hardware through the \textit{AtomicSimpleCPU} model available in \textit{gem5} configured to emulate the x86 \ac{ISA}. The software stack included a Linux kernel and a set of target tasks of different complexity. Six MiBench \cite{Guthaus:2001aa} applications were considered: \texttt{qsort}, \texttt{dijkstra}, \texttt{susan}, \texttt{sha}, \texttt{bitcount} and \texttt{basicmath}. 
The fault injection task was accomplished using \textit{FIMSIM}, an add-on to gem5 that enables fault injections of permanent (stuck-at) faults, \ac{SBU}, and \ac{MBU} in different microprocessor structures. This preliminary study focused on \ac{SBU} in the \ac{intRF}. After the fast-forward cycle required to boot the operating system, faults were injected at random locations and time intervals during the execution of the application. \textit{gem5} enables monitoring of the internal microprocessor state using \textit{checkpoints} (i.e., snapshots of the hardware architecture containing all the inner values at a particular clock tick) and \textit{stats} (i.e., performance counters that profile the number of internal events, such as cache misses, jumps, accesses to memory, and so on). To speed up the experiments, every fault injection run started from a checkpoint collected at the end of the fast-forward cycle corresponding to the end of the Linux boot process (CPs). Several \textit{stats} (S1, S2, ..., Sn) were collected during the simulation to consider the time dimension when building the final dataset. 

The fault effect was classified by comparing the last checkpoint of the fault injection run (CPn) with the golden execution (CPe). Possible outcomes following a fault injection can be explained in five categories: (i) \textit{Crash}: the program completely stops working and exits; (ii) \textit{\ac{SDC}}: the program terminates, but its outcome is wrong; (iii) \textit{Benign}: even if there was a fault, the program's outcome is correct (iv) \textit{Hangs}: the program is stuck within a loop; (v) \textit{Reboots}: the operating system reboots. Crashes and hangs are straightforward to detect by inspecting the program counter \cite{Ragel:2008}, so they are not considered in this study. Reboots cannot instead be traced with the available simulation setup. Hence, this study focuses on detectors able to discriminate \ac{SDC} and Benign executions. 

The \ac{HPC} in the \textit{gem5 stats} collected during fault injection experiments were used to create the dataset analyzed in the next section. A total amount of about 600 features were monitored during every fault injection run. Data were preprocessed and normalized before carrying out the analysis. First, some values in the collected dataset were non-numerical (NaN). Features with more than 5\% NaN values and attributes with zero variance across experiments were removed. At this point, simulations containing even a single NaN value were removed from the dataset. Finally, missing values were set to zero, and the dataset was normalized. \ac{SDC} and Benign classes are not balanced in a fault injection experiment. To enable a fair analysis, the dataset was balanced using downsampling on the Benign (major) class. \autoref{tab:dataset} summarizes the characteristics of the final dataset after preprocessing.

\begin{table*}[!hbt]
  \caption{Summary of the dataset structure.}
  \label{tab:dataset}
  \centering
  \begin{tabular}{|l|cccccc|}
  	\hline
    Benchmark 	& \#Simulations  & \#Features  & \#(Giga) Ticks  & \% Benign & \% SDC  &  \% crash/hang \\
    \hline
    qsort 		& 25,000 	& 366 	& 86.5 	& 72	.1	& 	25.6		&	2.3	\\
    dijkstra 	& 29,180 	& 378 	& 64.2 	& 79	.2	& 	20.2 	&	0.6	\\
    susan 		& 21,440 	& 363 	& 35.1 	& 67.9	& 	26.5		&	5.6	\\
    sha 		& 13,809 	& 359 	& 18.5 	& 72.6		& 	27.2		&	0.2	\\
    bitcount 	& 25,000 	& 360 	& 9.8  	& 81	.2	& 15.6		&	3.2	\\
    basicmath 	& 25,000 	& 358 	& 6.5 	& 92.7	& 6.4 		&	0.9	\\
    \hline
  \end{tabular}
\end{table*}

Since part of the exploration aims to understand if the temporal dimension adds any information to the collected data, increasing the fault detection accuracy, the execution of the different benchmarks was profiled over time by collecting \textit{gem5 stats} with different granularities, i.e., every 10, 20, 50, and 100 simulation ticks, where the total ticks within a benchmark are given in \autoref{tab:dataset}. The table also reports, for each benchmark, the features included in the model investigation, the number of ticks for executing it (expressed in Giga), and the percentage of Benign, \ac{SDC} and crash and hangs classifications among the faulty runs.

%% file: Sections/Results/results.tex
\section{Results}
\label{sec:results}

	This section reports a comprehensive analysis of the collected dataset to try to answer the question proposed at the beginning of this paper, i.e., if microarchitectural features are sufficient to build machine learning models able to detect soft errors in microprocessor-based systems.
	 
	\input{Sections/Results/Subsections/dataAnalysis}
	\input{Sections/Results/Subsections/mlModels}

%% file: Sections/Results/Subsections/dataAnalysis.tex
\subsection{Data analysis}
\label{subsec:dataAnalysis}

\begin{figure*}[!t]
			\centering
			\subfigure[qSort full dataset]{\includegraphics[trim={1.2cm 1.2cm 1.2cm 1.2cm},clip,width=0.15\textwidth] {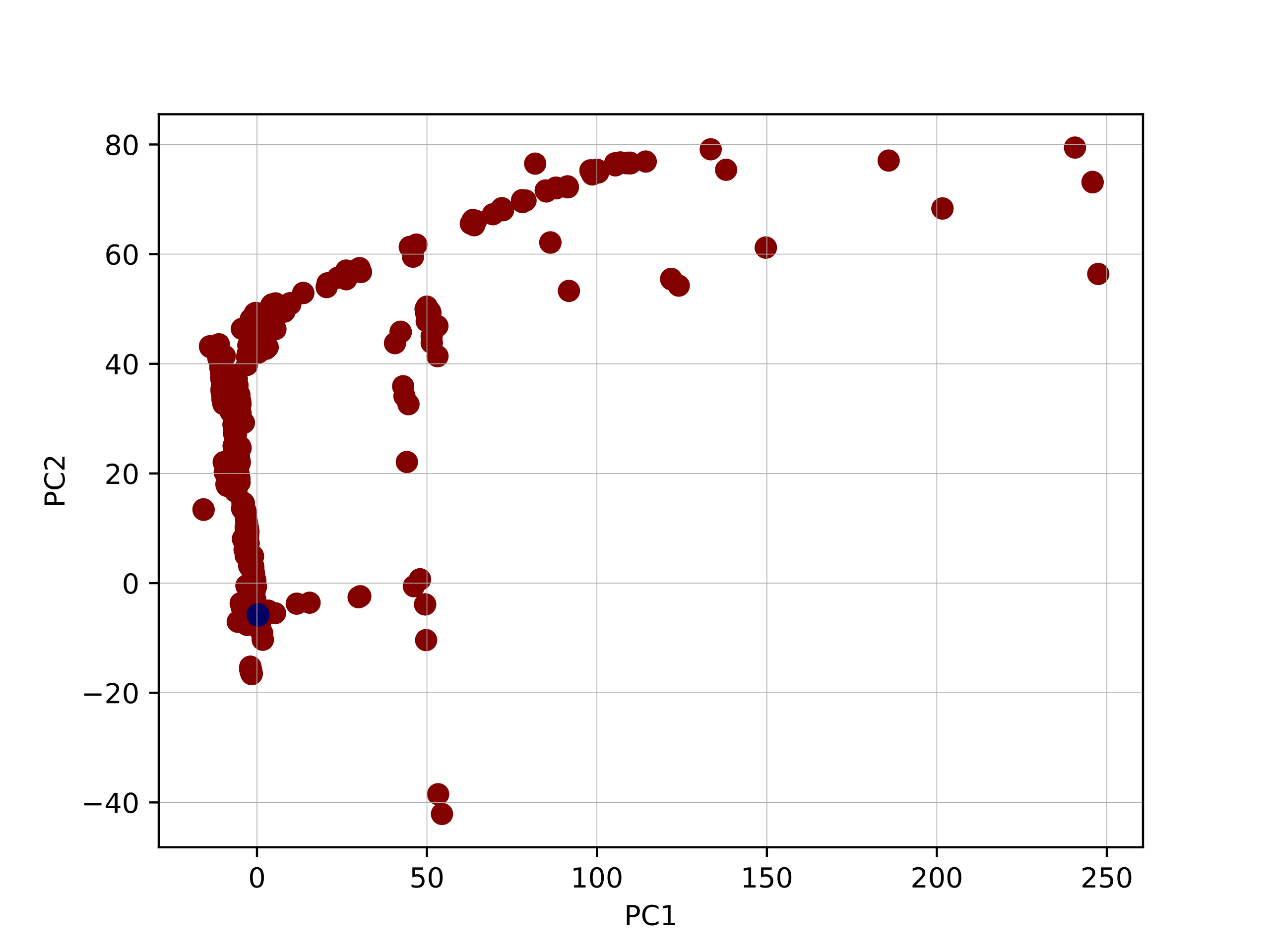} \label{fig:pcaQsort}}
			\subfigure[Dijkstra full dataset]{\includegraphics[trim={1.2cm 1.2cm 1.2cm 1.2cm},clip,width=0.15\textwidth] {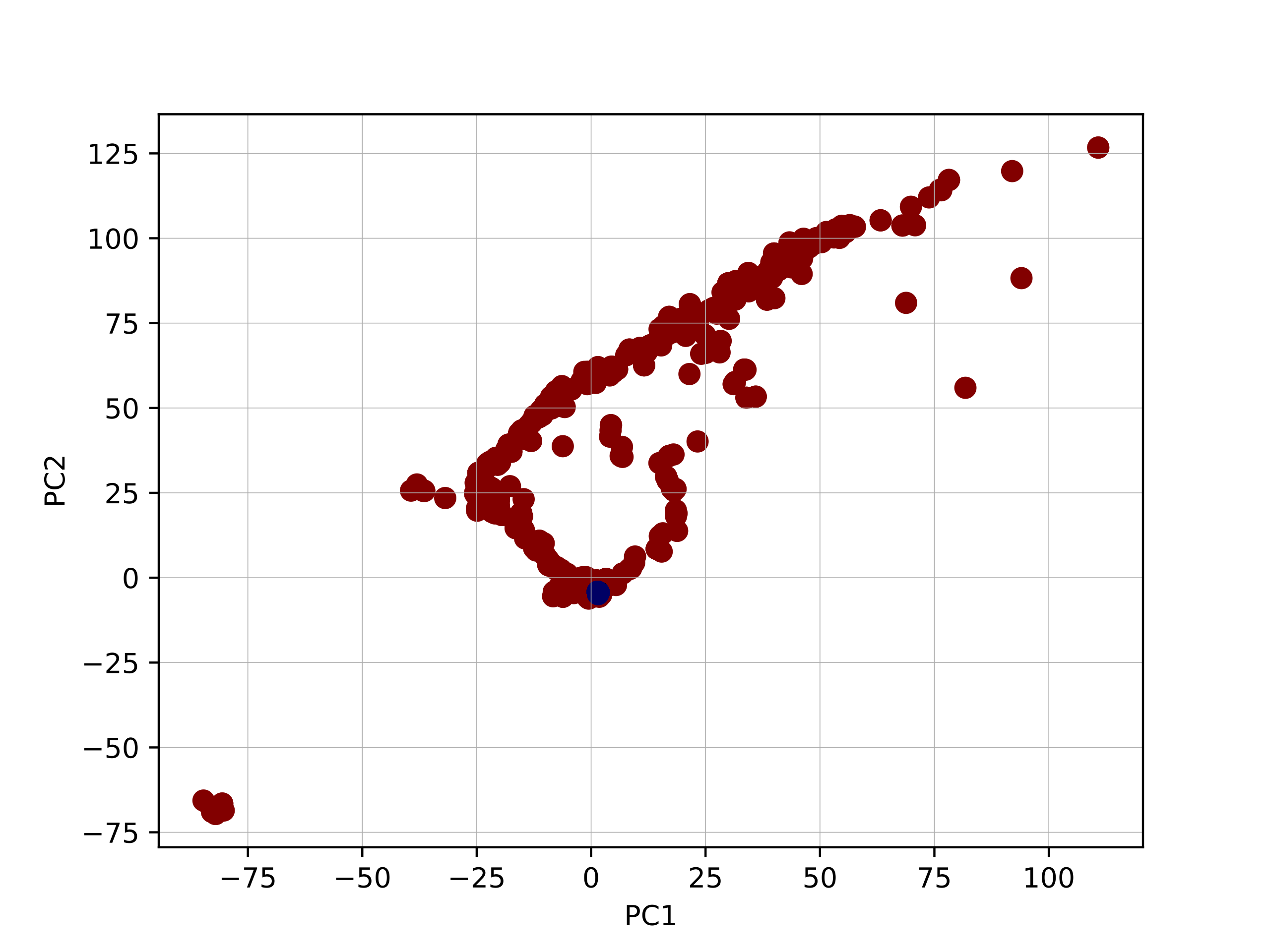} \label{fig:pcaDijkstra}}
			\subfigure[Susan full dataset]{\includegraphics[trim={1.2cm 1.2cm 1.2cm 1.2cm},clip,width=0.15\textwidth] {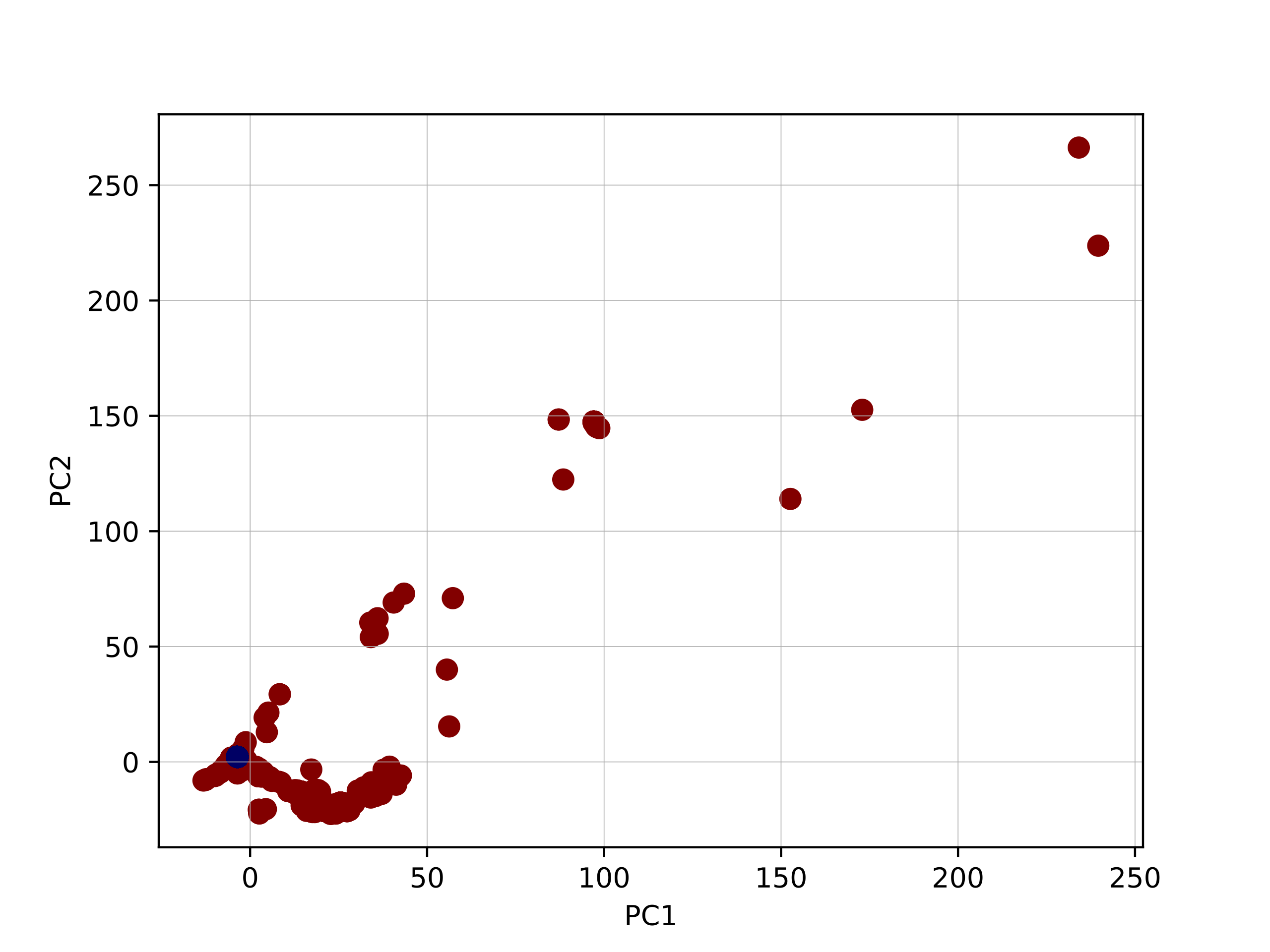} \label{fig:pcaSusan}}
			\subfigure[Sha full dataset]{\includegraphics[trim={1.2cm 1.2cm 1.2cm 1.2cm},clip,width=0.15\textwidth] {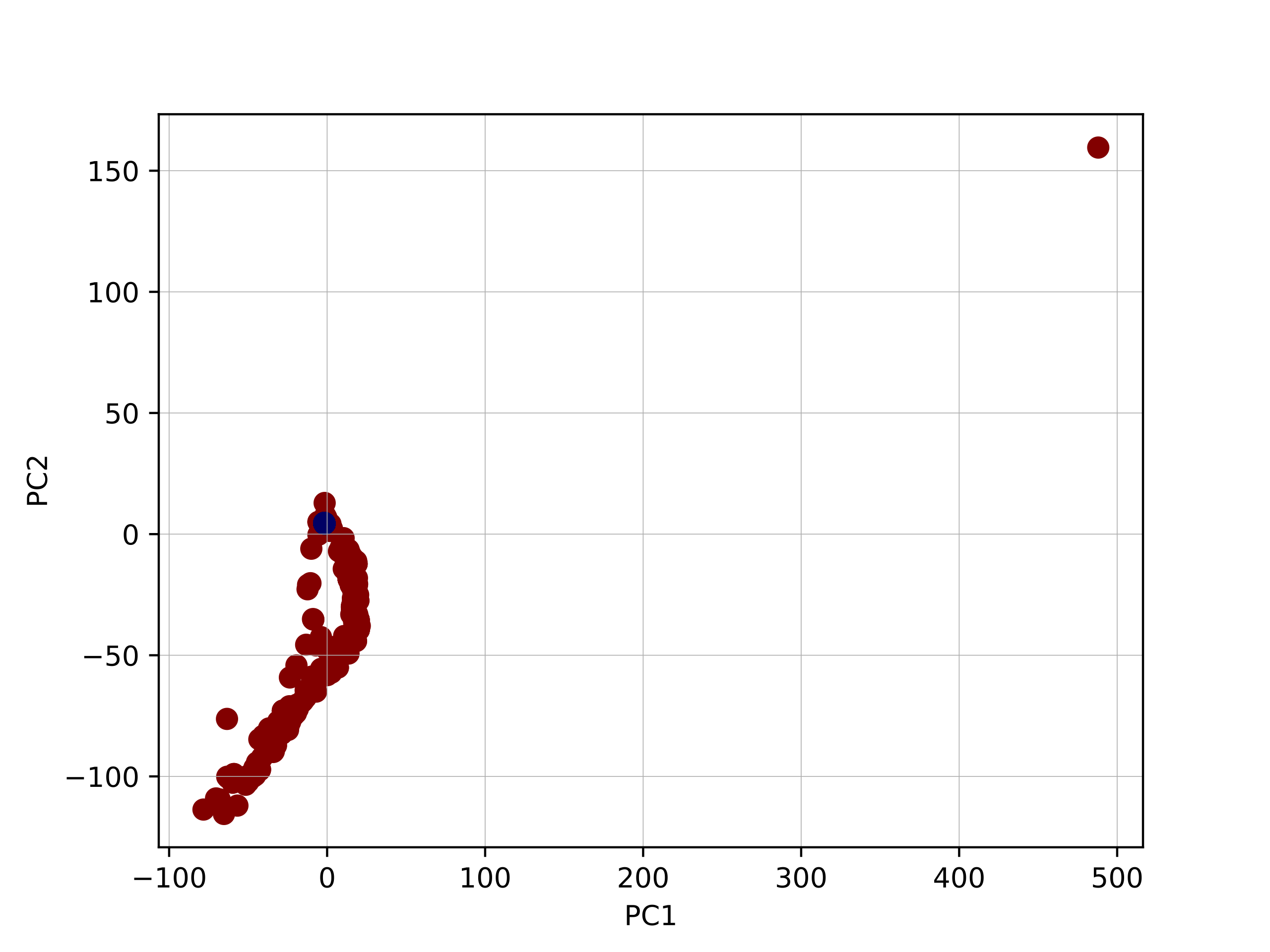} \label{fig:pcaSha}}
			\subfigure[bitcount full dataset]{\includegraphics[trim={1.2cm 1.2cm 1.2cm 1.2cm},clip,width=0.15\textwidth] {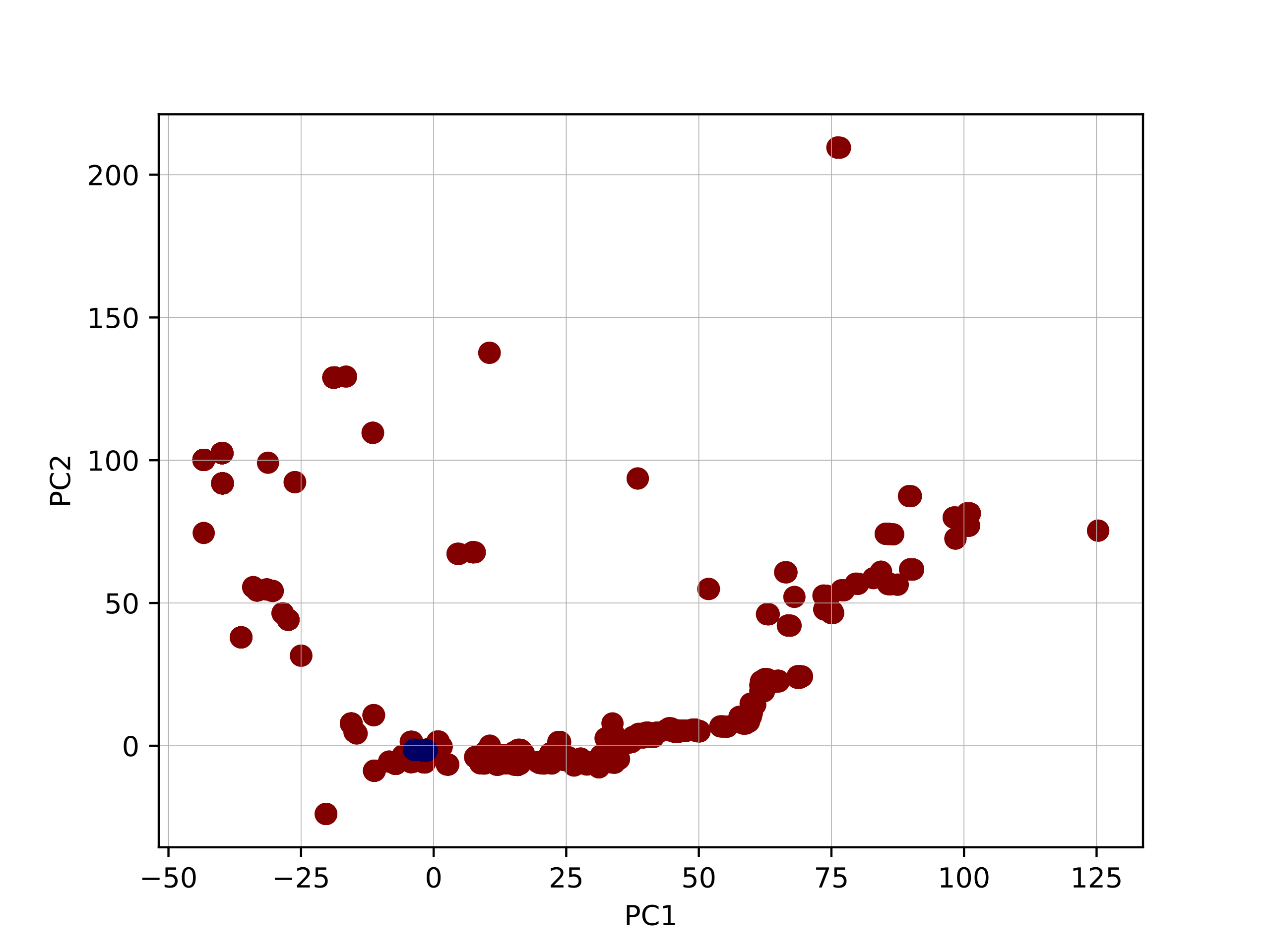}\label{fig:pcaBit}	}
			\subfigure[basicmath full dataset] {\includegraphics[trim={1.2cm 1.2cm 1.2cm 1.2cm},clip,width=0.15\textwidth] {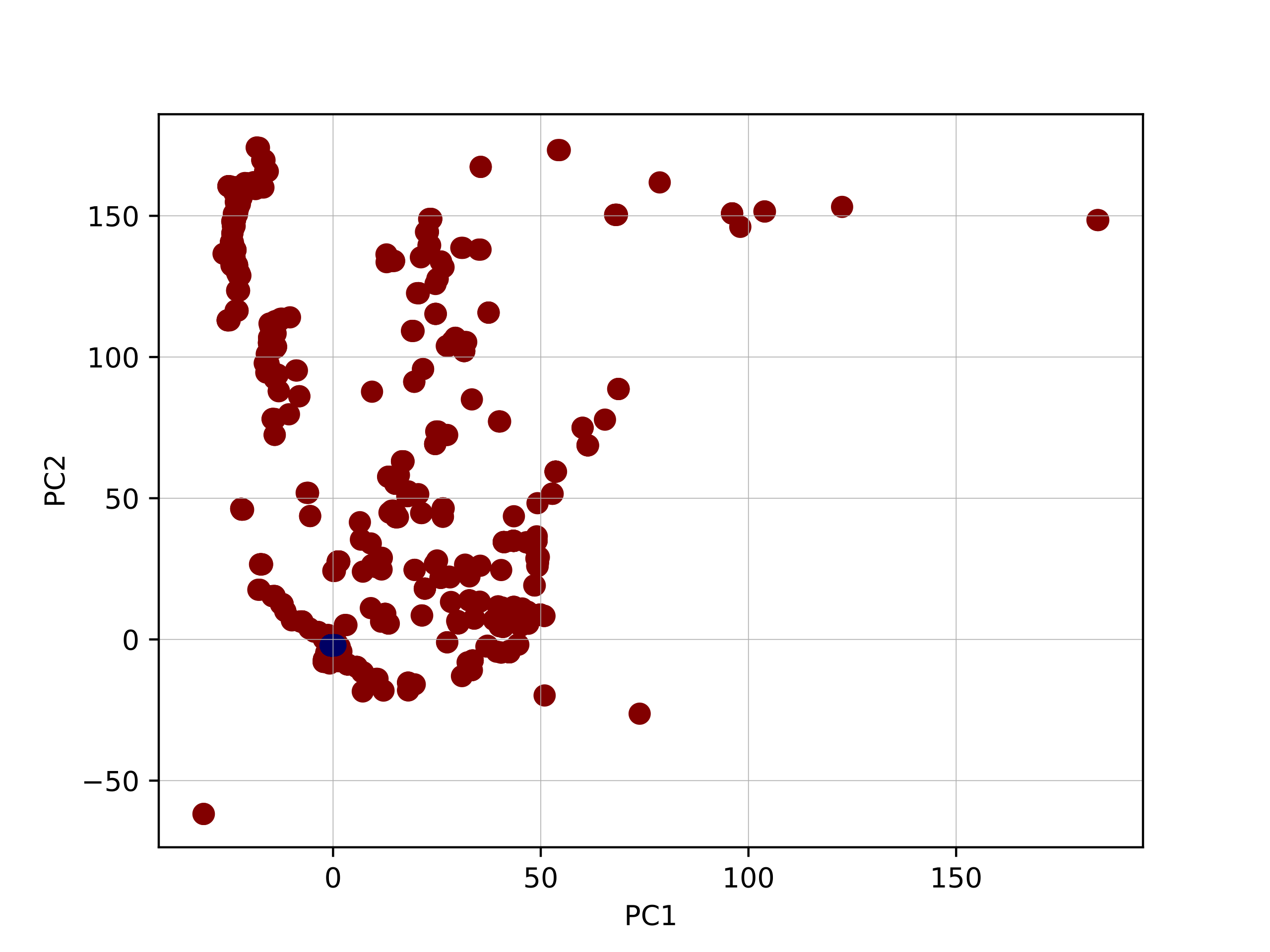} \label{fig:pcaMath}}
			
			\subfigure[qSort \acs{HTDR}]{\includegraphics[trim={1.2cm 1.2cm 1.2cm 1.2cm},clip,width=0.15\textwidth] {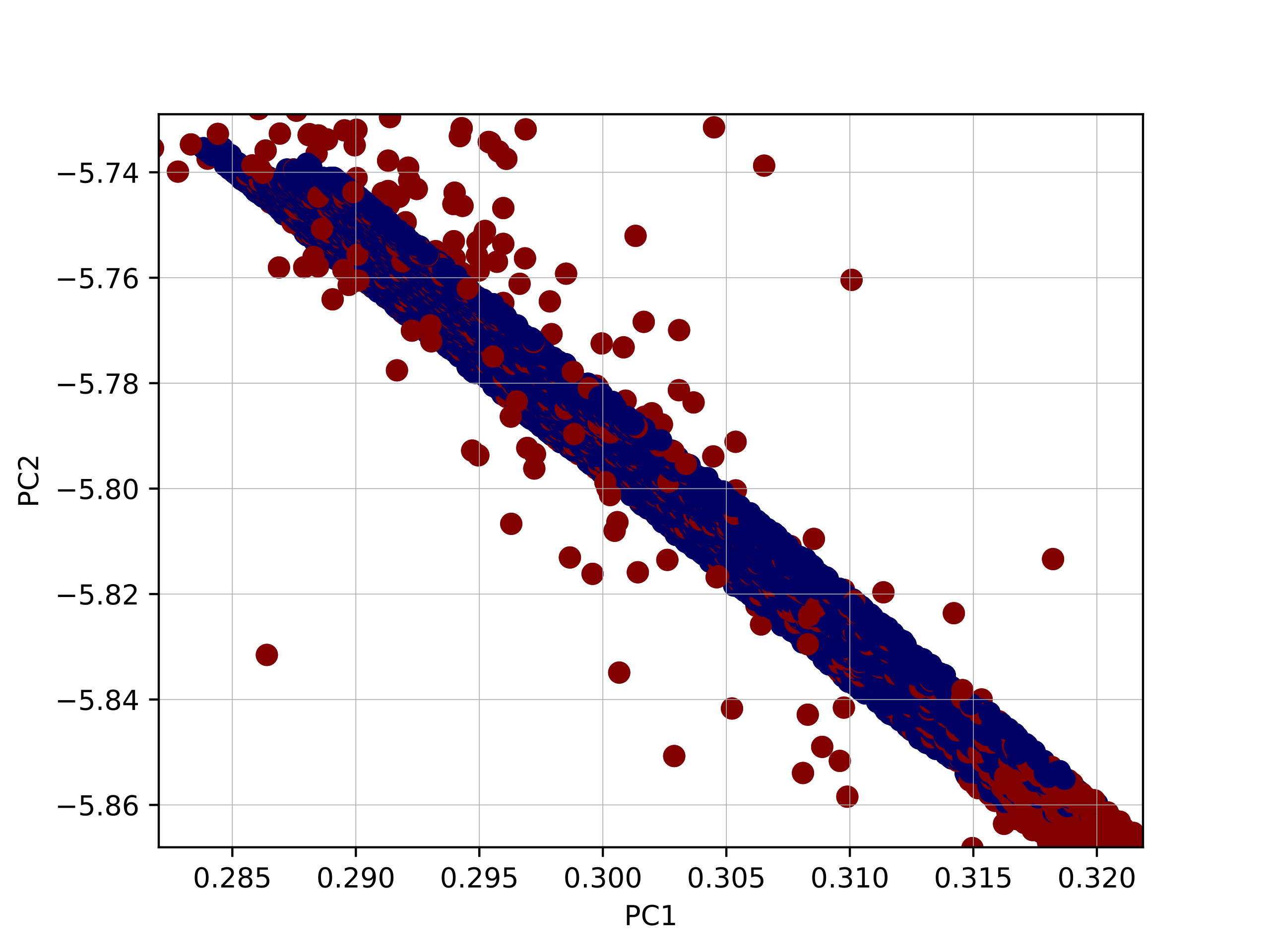} \label{fig:pcaQsort_z}}
			\subfigure[Dijkstra \acs{HTDR}]{\includegraphics[trim={1.2cm 1.2cm 1.2cm 1.2cm},clip,width=0.15\textwidth] {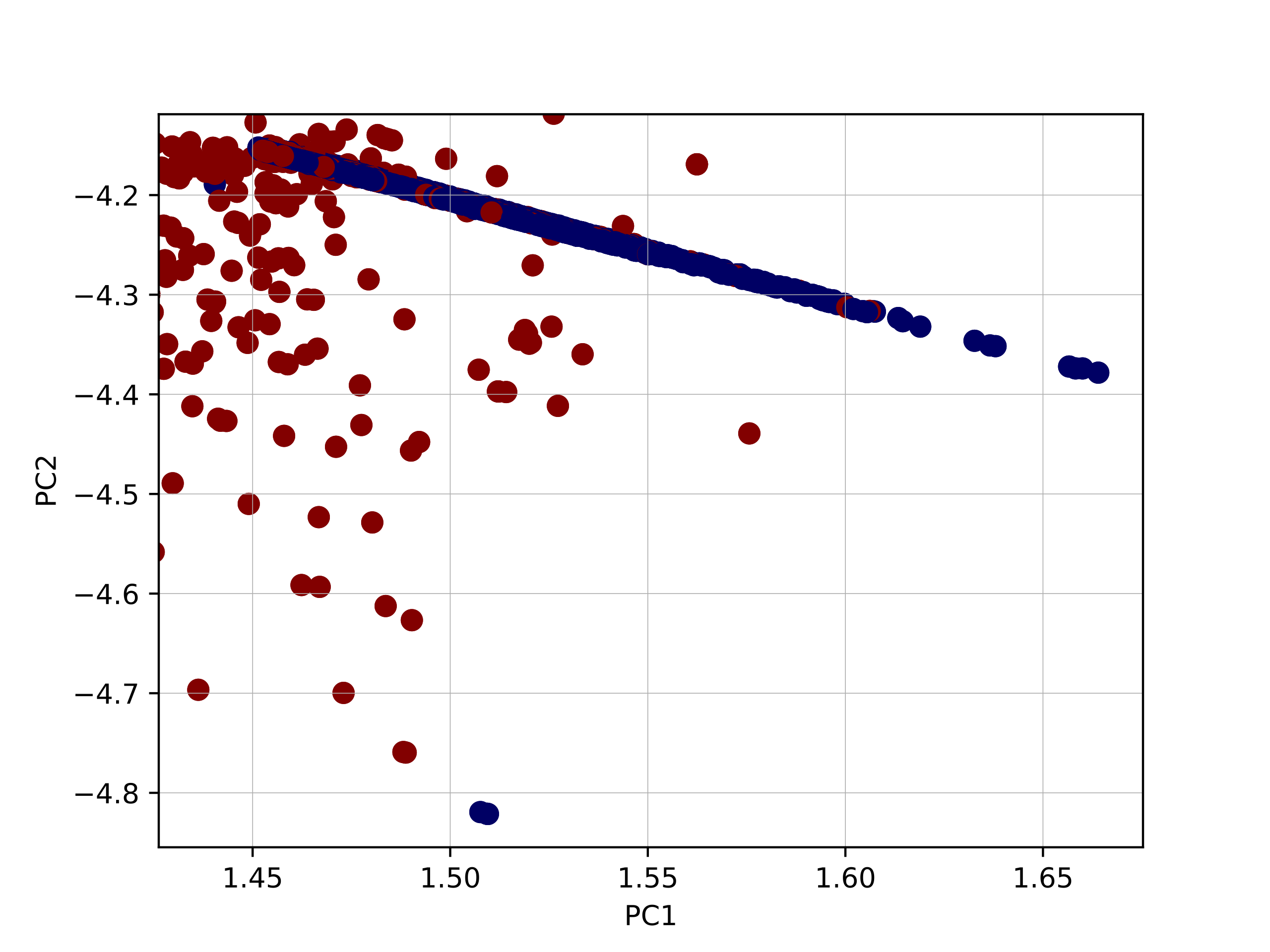} \label{fig:pcaDijkstra_z}}
			\subfigure[Susan \acs{HTDR}]{\includegraphics[trim={1.2cm 1.2cm 1.2cm 1.2cm},clip,width=0.15\textwidth] {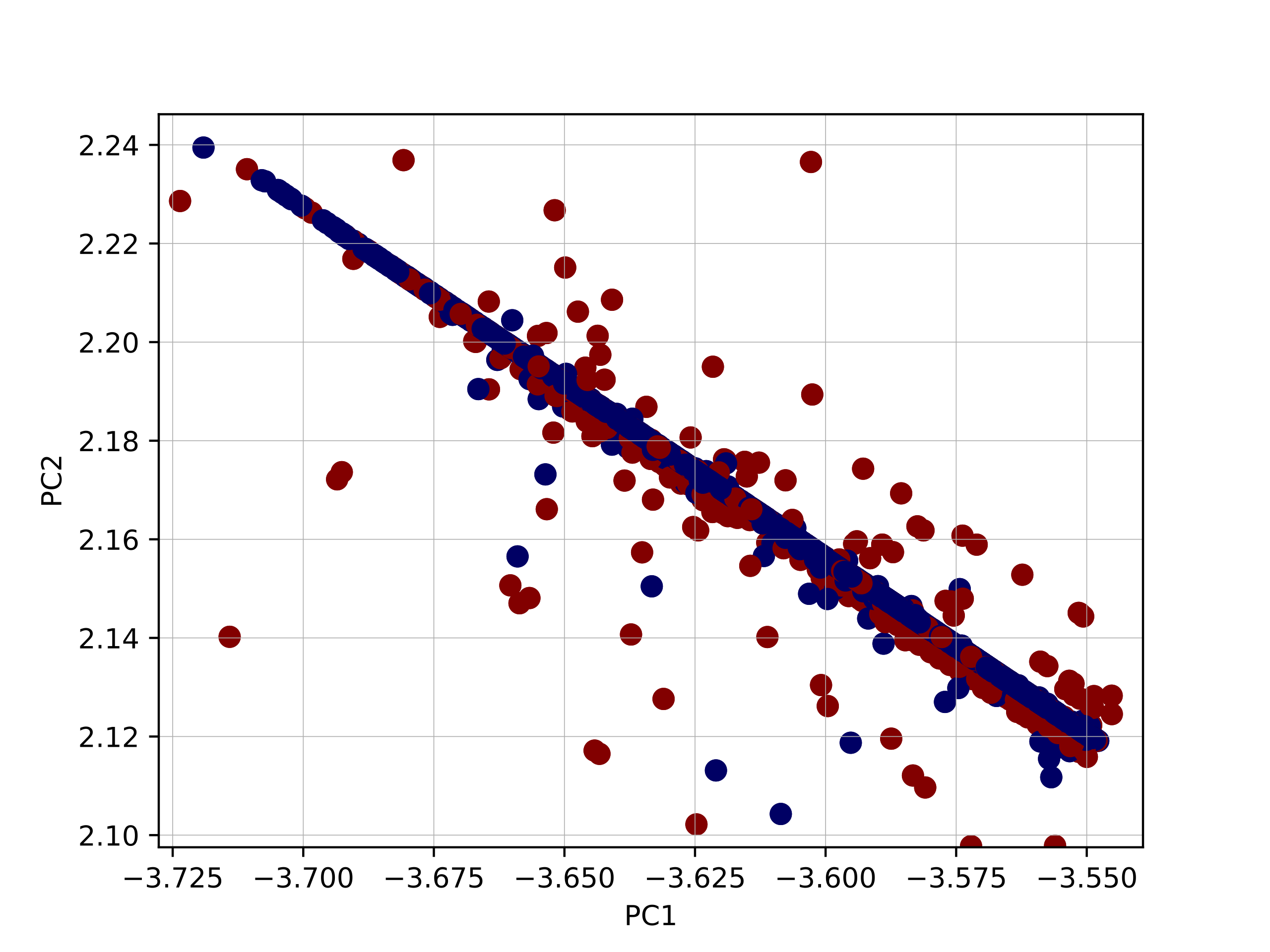} \label{fig:pcaSusan_z}}
			\subfigure[Sha \acs{HTDR}]{\includegraphics[trim={1.2cm 1.2cm 1.2cm 1.2cm},clip,width=0.15\textwidth] {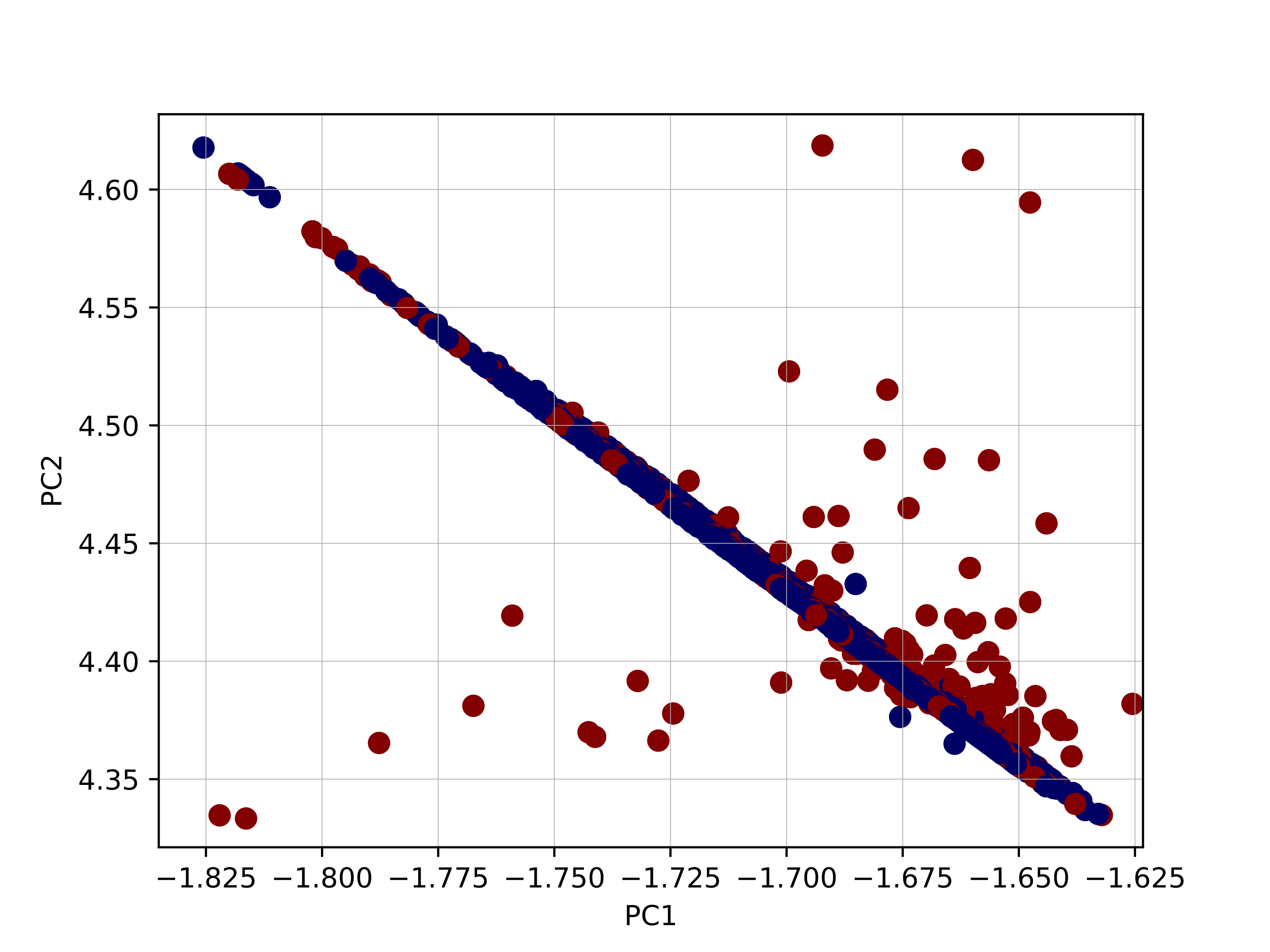} \label{fig:pcaSha_z}}
			\subfigure[bitcount \acs{HTDR}]{\includegraphics[trim={1.2cm 1.2cm 1.2cm 1.2cm},clip,width=0.15\textwidth] {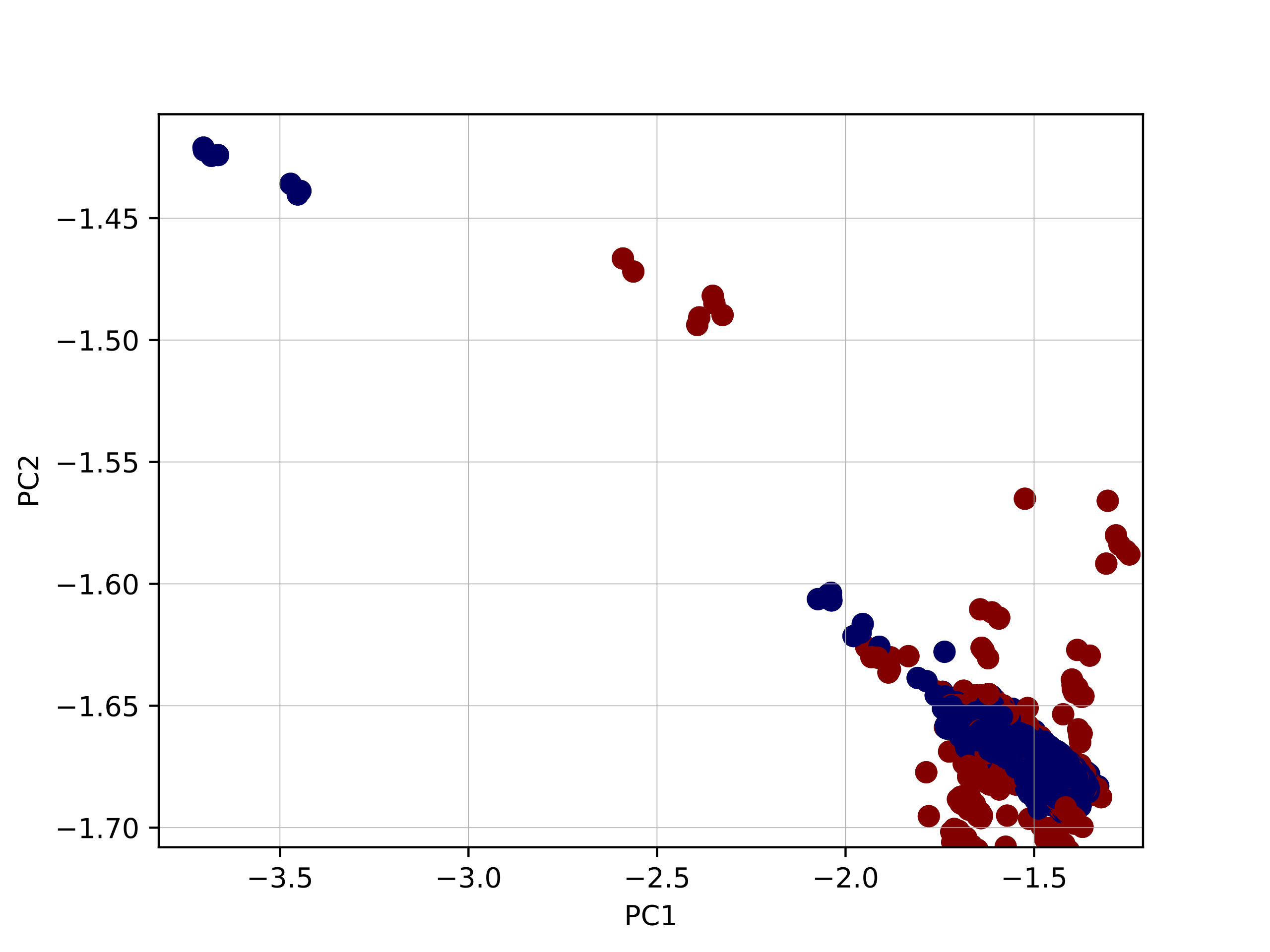}\label{fig:pcaBit_z}	}
			\subfigure[basicmath \acs{HTDR}] {\includegraphics[trim={1.2cm 1.2cm 1.2cm 1.2cm},clip,width=0.15\textwidth] {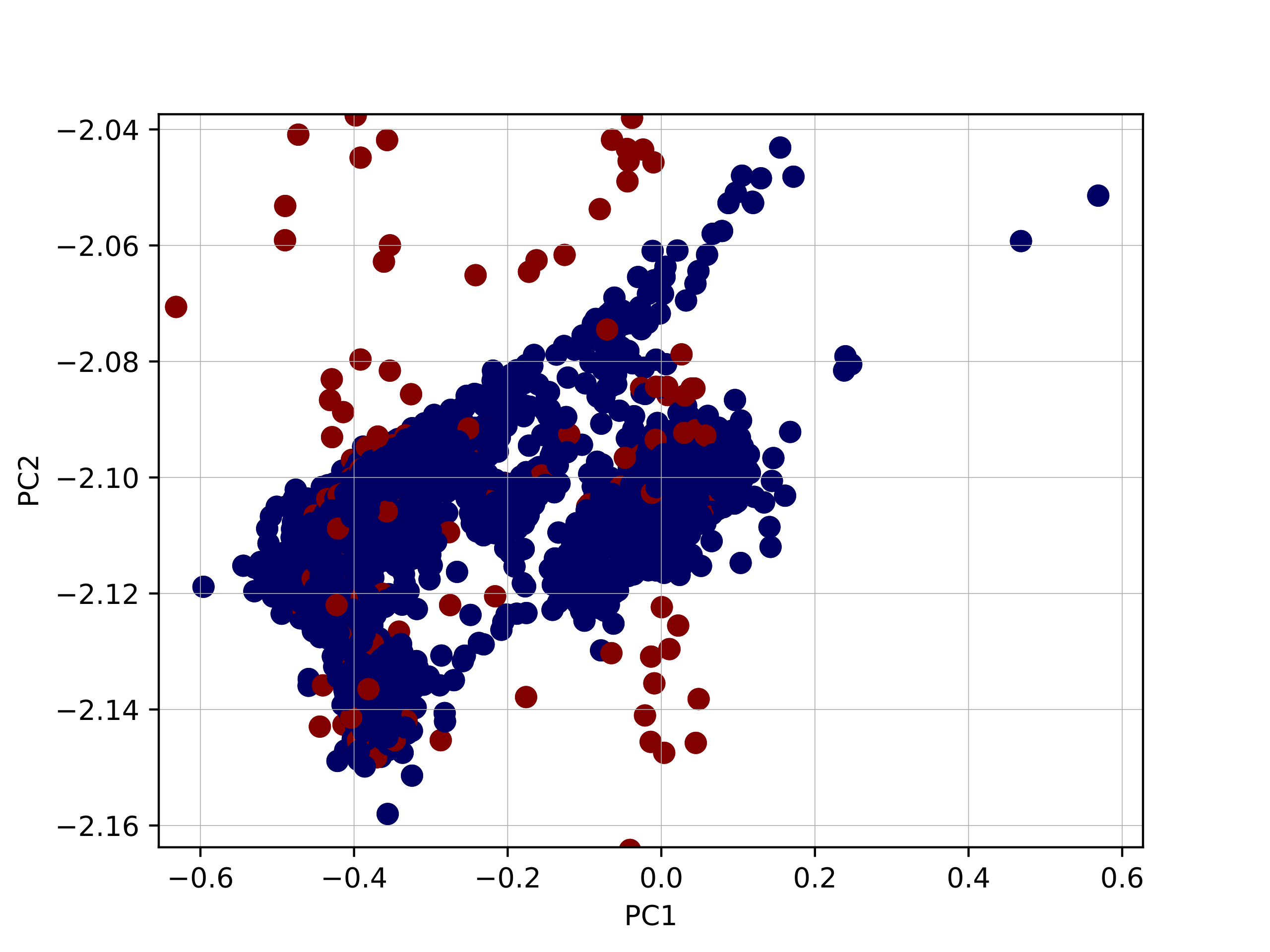} \label{fig:pcaMath_z}}
				
			\caption{\small{Visual representation of the dataset for the six benchmarks using PCA. It shows faulty runs (red) and benign runs (blue)}. Subfigures (a-f) show the full dataset, while (g-l) show the \acf{HTDR}}
			\label{fig:pcaFull}
		\end{figure*}

 The first goal of the analysis is a preliminary inspection of the quality of the selected features. For this purpose, we used Principal Component Analysis (PCA). \autoref{fig:pcaFull} provides a visual representation of the datasets for the six benchmarks plotting the first two principal components. 

When analyzing the complete datasets (Figures \ref{fig:pcaQsort}, \ref{fig:pcaDijkstra}, \ref{fig:pcaSusan}, \ref{fig:pcaSha}, \ref{fig:pcaBit}, and \ref{fig:pcaMath}), benign executions (blue dots) cluster in a small portion of the plot, while faulty executions (red dots) instead scatter over the plot, this suggests that the corruption of a single bit generates a significant deviation in the microarchitecture's internal features, resulting in cases that can be detectable from a reliability standpoint. However, Figures \ref{fig:pcaQsort_z}, \ref{fig:pcaDijkstra_z}, \ref{fig:pcaSusan_z}, \ref{fig:pcaSha_z}, \ref{fig:pcaBit_z}, and \ref{fig:pcaMath_z} highlight the presence in all datasets of a \acf{HTDR} of overlapping samples of the two classes. Since \ac{AI} models are known to struggle with such data distributions, these regions must be carefully considered to avoid a significant loss in accuracy.

\begin{figure}[!ht]
			\centering
			\includegraphics[trim={0cm 0cm 0cm 0cm},clip,width=\columnwidth]
			{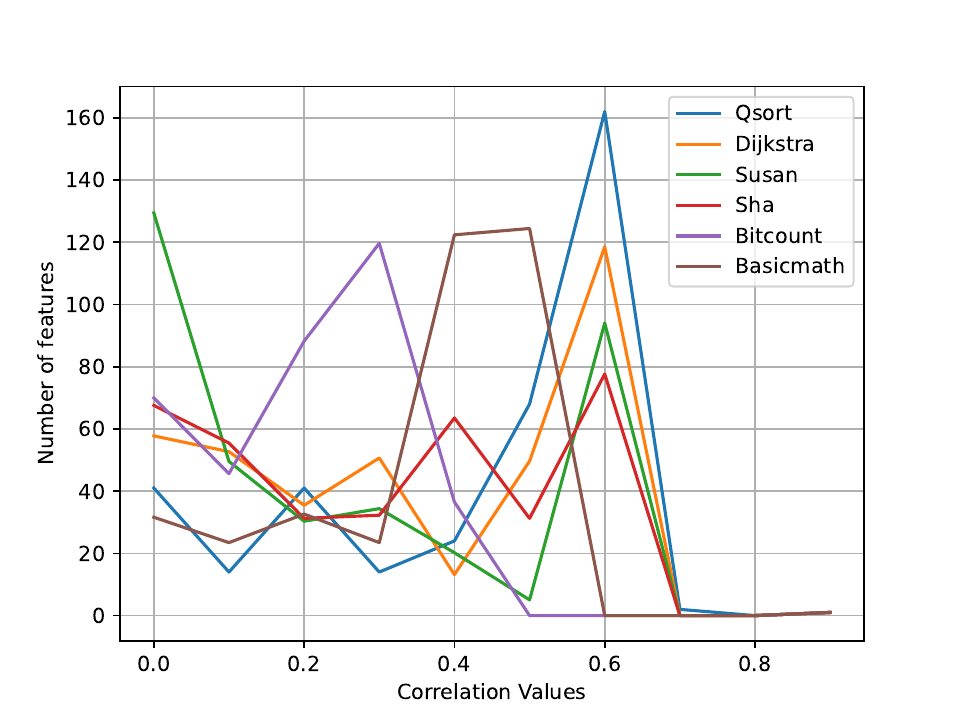}
			\caption{Statistical distribution of the correlation
			values between features and fault classification for all six benchmarks}
			\label{fig:corr}
		\end{figure}

Up to now, the analysis considered all the features in the dataset, which are a considerable number, even after the data cleaning. Therefore, analyzing the correlation of each attribute with the final fault classification can guide through a feature selection phase. Pearson’s correlation coefficient was used to perform this analysis due to its minor sensitivity to false positives. \autoref{fig:corr} shows the distribution of the correlation values for the six benchmarks.
The analysis highlights different results depending on the benchmark. In the case of \texttt{qsort}, \texttt{dijkstra}, \texttt{susan}, and \texttt{sha}, most features correlate more than 0.5 with faulty executions. 
This means that many microarchitectural features can be used to detect a faulty run. Results for \texttt{basicmath} are weaker but still acceptable, while \texttt{bitcount} reports a low correlation. This aspect will be better investigated in the next section.  

\begin{figure*}
			\centering
				\subfigure[All features]{\includegraphics[width=0.85\textwidth] {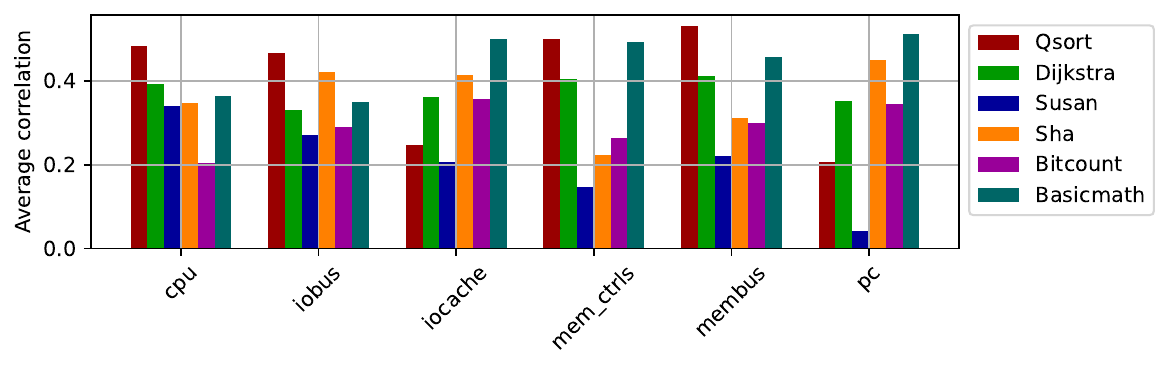} \label{fig:secondaryFeaturesALL}}
				\subfigure[CPU related features]{\includegraphics[width=0.85\textwidth] {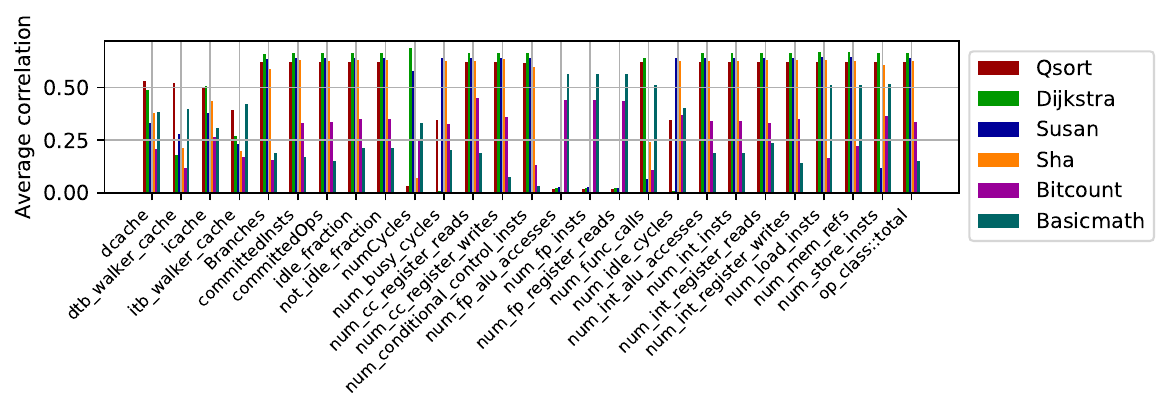} \label{fig:secondaryFeaturesCPU}}
	\caption{Statistical distribution of the correlation values over the different \textit{gem5} classes for the six benchmarks.}
	\label{fig:secondaryFeatures}
			\end{figure*}

Moving to the microarchitectural level, following the \textit{gem5} hierarchical organization of features into six main sub-classes, \autoref{fig:secondaryFeatures} shows the correlation of each group with faulty executions. Each of the six main sub-classes of attributes shows differences depending on the benchmarks (Figure \autoref{fig:secondaryFeaturesALL}). However, when the correlation is significant, few features emerge as indicators of faulty execution, such as the ones related to memory accesses (\textit{mem\_ctrls} and \textit{membus}), to the input/output bus and the CPU. This last class includes many internal features, so it is worth analyzing them in more detail. 

Figure \autoref{fig:secondaryFeaturesCPU} reports the CPU class features split into its main sub-classes, showing a higher variability in attribute correlation across different benchmarks. Looking at \texttt{qsort}, the benchmarks with higher correlation, the sub-classes mostly affected are accessed to both data and instruction caches, meaning that faults impact the instruction fetch order and the access to data. Follows the number of branches, an indication that the execution flow has been modified, and the committed instructions and operations, evidence of what has been seen with the cache accesses. Eventually, deviations are also evident in the number of function calls, the internal registers' read and write operations, and the main memory accesses for load and store operations. 
This analysis suggests that when single bit-flips significantly impact the program execution flow, the microarchitectural features deviate from average, allowing detection. This important characteristic will be further discussed when considering using these datasets with dedicated machine-learning models. 

Finally, the dataset was analyzed considering the temporal dimension by sampling all features at different simulation times. The observed behavior is consistent with the one observed in the previous cumulative case. Interestingly, fault executions can still be detected, and \ac{HTDR} areas are present, suggesting that using microarchitectural features, it is feasible to start the detection early during the execution instead of waiting for the program to end.


To summarize, this preliminary analysis allowed us to draw some initial insights. A reasonably simple model is expected to provide good fault detection capabilities for the data that can be separated. At the same time, pure microarchitectural attributes might prove insufficient due to \ac{HTDR}. 
Including the temporal dimension to the data is not improving the fault detection but confirms that the detection can happen during the execution and not only at the end.

%% file: Sections/Results/Subsections/mlModels.tex
\subsection{Machine learning models}
\label{subsec:mlModels}

Starting from the preliminary analysis proposed in \autoref{subsec:mlModels}, this section studies how different machine learning models can be trained to detect soft errors based on the considered attributes, highlighting strengths and weaknesses. For this analysis, the datasets were split as follows: 60\% for the training set, 15\% for the validation set, and 25\% for the test set. By using correlations (see \autoref{fig:corr}), we were able to identify the following collection of 19 features consistently highly correlated in most benchmarks (the reader may refer to the gem5 documentation for a detailed description of each feature): 

	{\scriptsize
\begin{verbatim}
1.  system.cpu.dtb_walker_cache.tags.age_task_id_blocks_1024::2
2.  system.cpu.icache.tags.age_task_id_blocks_1024::2
3.  system.membus.pkt_count_system.apicbridge.master::total
4.  system.membus.trans_dist::MessageResp
5.  system.iobus.pkt_size_system.pc.south_bridge.io_apic.
int_master::total
6.  system.iobus.pkt_count_system.pc.south_bridge.io_apic.
int_master::system.apicbridge.slave
7.  system.membus.trans_dist::MessageReq
8.  system.membus.pkt_size_system.apicbridge.master::system.
cpu.interrupts.int_slave
9.  system.membus.pkt_count_system.apicbridge.master::system.
cpu.interrupts.int_slave
10. system.iobus.pkt_size_system.pc.south_bridge.io_apic.
int_master::system.apicbridge.slave
11. system.iobus.trans_dist::MessageResp
12. system.membus.pkt_size_system.apicbridge.master::total
13. system.iobus.pkt_count_system.pc.south_bridge.io_apic
.int_master::total
14. system.iobus.trans_dist::MessageReq
15. system.membus.pkt_size_system.cpu.dcache.mem_side::system
.cpu.interrupts.pio
16. system.membus.pkt_count_system.cpu.dcache.mem_side::system
.cpu.interrupts.pio
17. system.cpu.dcache.tags.total_refs
18. system.iocache.demand_miss_rate::total
19. system.iocache.ReadReq_miss_rate::total
\end{verbatim}}

They were used to demonstrate that a set of features can work with several benchmarks. All experiments used a 19-32-2 network architecture composed of three layers, including 19, 32, and 2 neurons. These numbers were tuned using trial and error.

The first model considered is the \ac{FC-FFNN}, trained using the cumulative dataset. 
		\begin{table}[h]
			\centering
		 \caption{\ac{FC-FFNN} performance metrics on the full dataset}
		 \label{tab:fcffnnFullDataset}
			\begin{tabular}{ |c|c|c|c| } 
				\hline
				Benchmark & Accuracy & F1 score \\ 
				\hline
				qsort 		& 91.10 & 79.91\\
				\hline
				dijkstra 	& 92.19 & 75.67\\
				\hline
				susan 		& 88.79 & 71.92\\
				\hline
				sha 			& 87.03 & 68.45\\
				\hline
				bitcount 	& 90.75 & 56.81\\
				\hline
				basicmath 	& 96.22 & 59.66\\
				\hline
			\end{tabular}
		\end{table}

\begin{figure*}[ht]
			\centering
		\subfigure[qsort]{\includegraphics[trim={1cm 0.0cm 1cm 0.5cm},clip,width=0.16\textwidth]
			{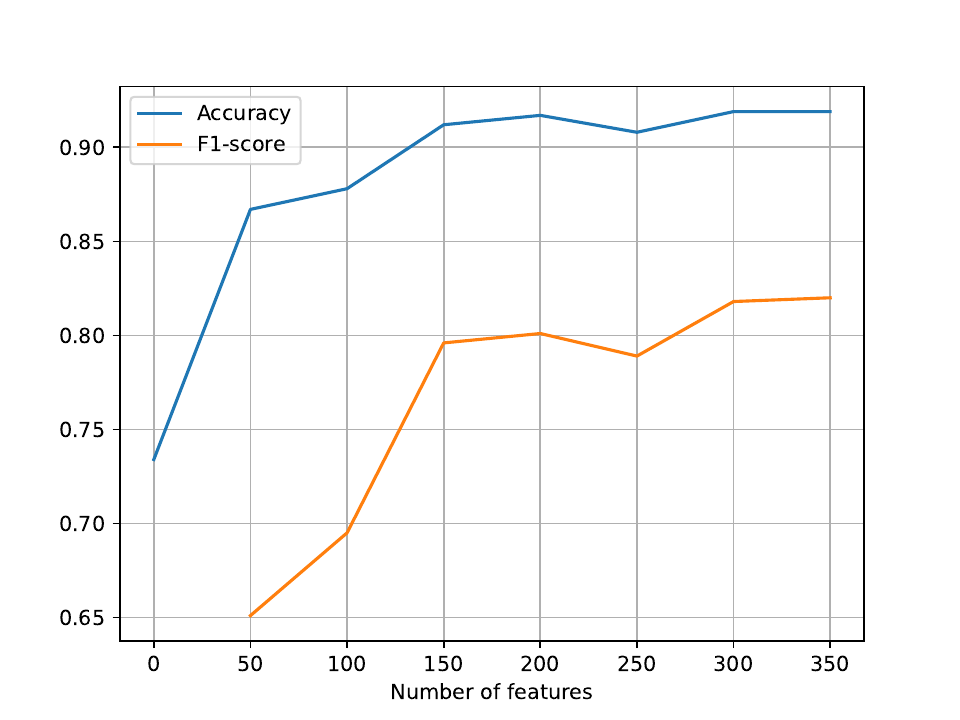}}
		\subfigure[djkstra]{\includegraphics[trim={1cm 0.0cm 1cm 1cm},clip,width=0.16\textwidth]
			{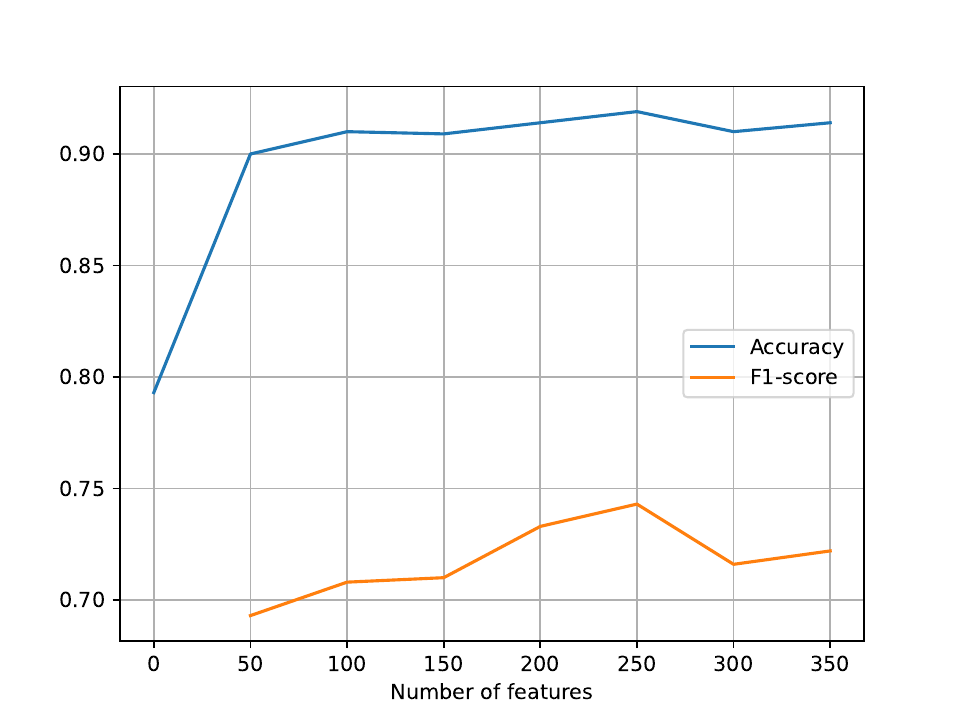}}			
		\subfigure[susan]{\includegraphics[trim={1cm 0.0cm 1cm 1cm},clip,width=0.16\textwidth]
			{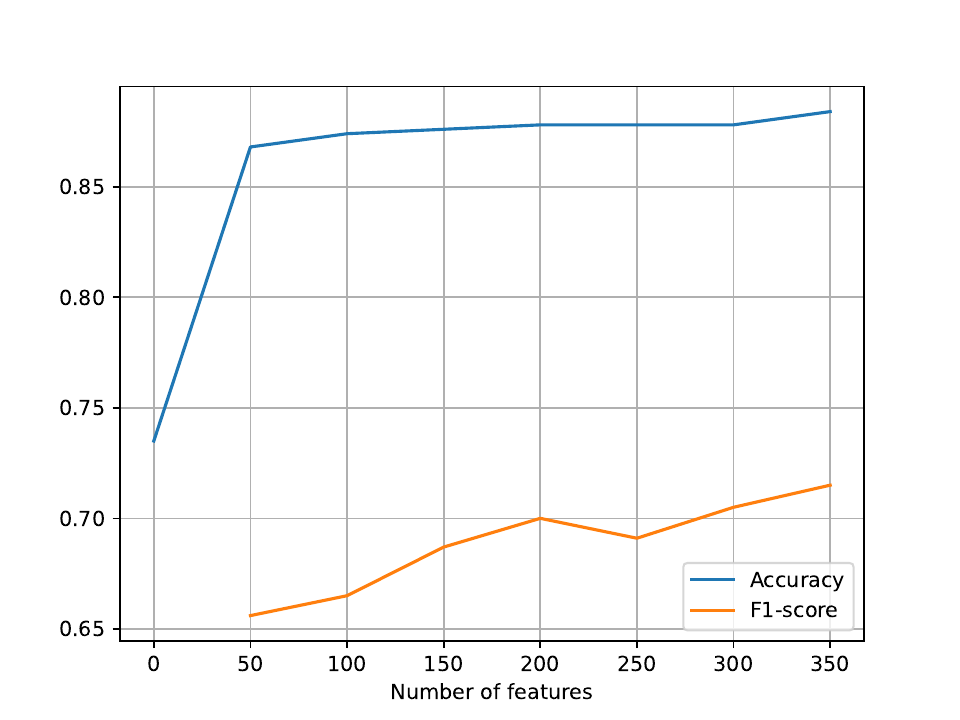}}	
		\subfigure[sha]{\includegraphics[trim={1cm 0.0cm 1cm 1cm},clip,width=0.16\textwidth]
			{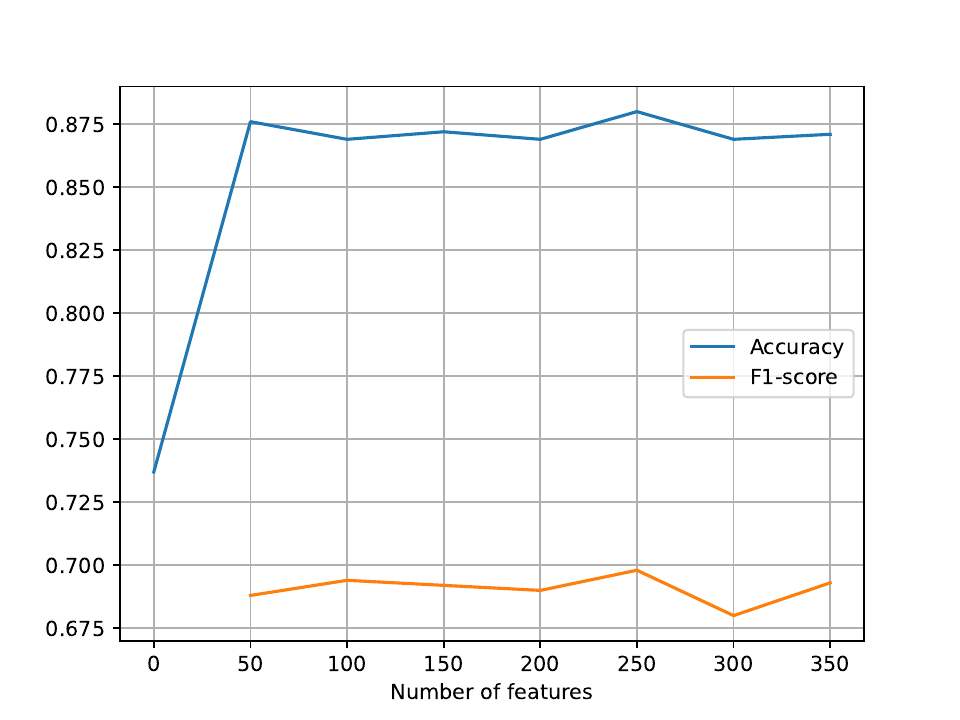}}
		\subfigure[bitcount]{\includegraphics[trim={1cm 0.0cm 1cm 1cm},clip,width=0.16\textwidth]
			{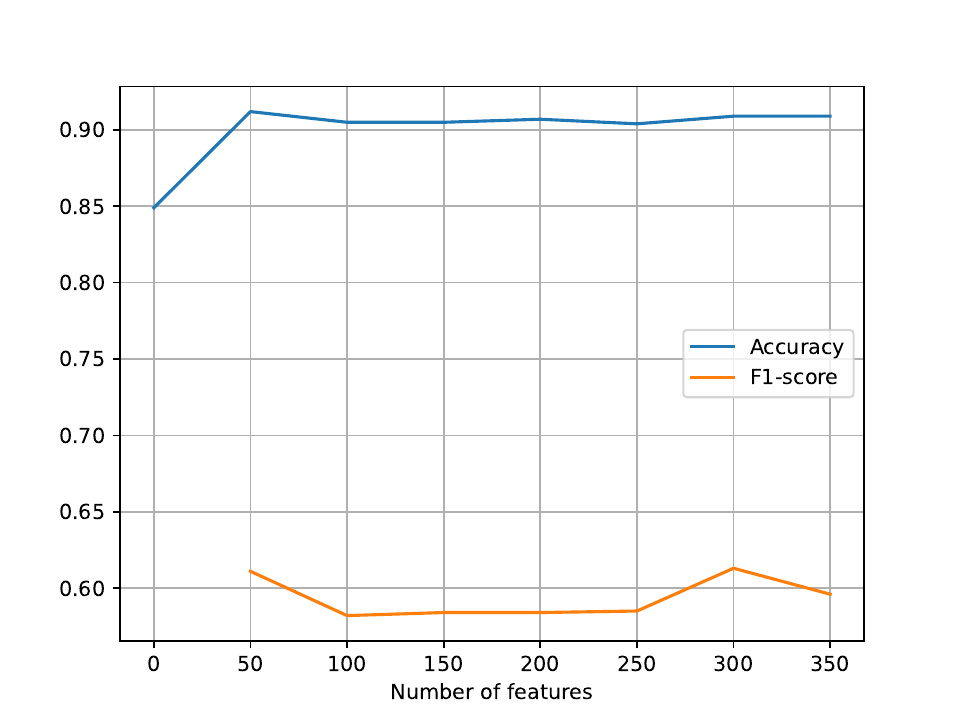}}
		\subfigure[basicmath]{\includegraphics[trim={1cm 0.5cm 1cm 1cm},clip,width=0.16\textwidth]
			{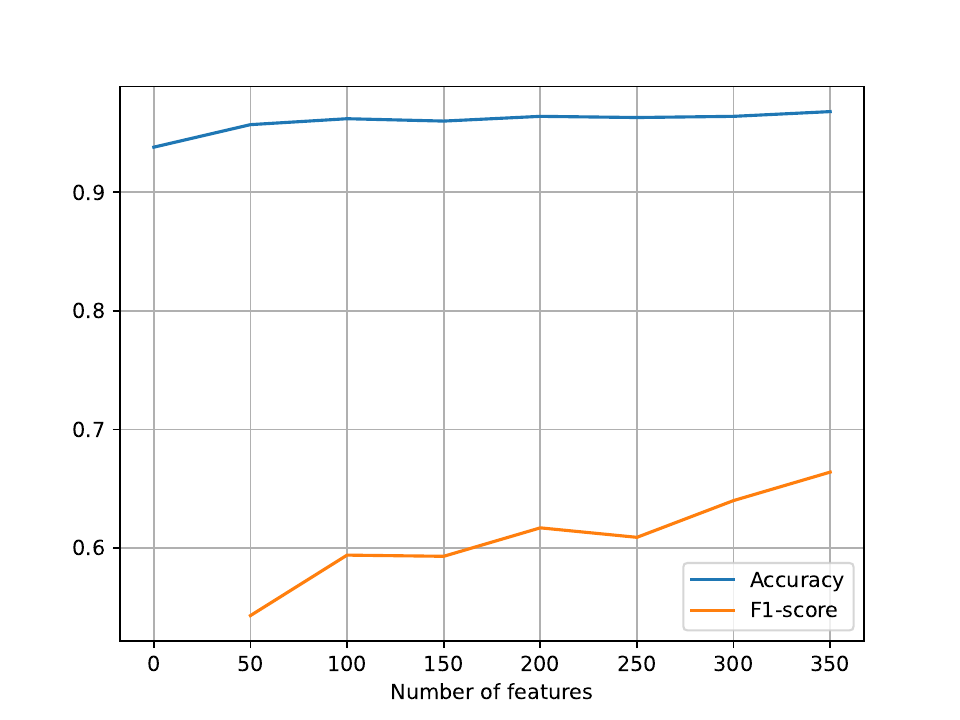}}
			
			\caption{Performance of several models trained on the full dataset with an increased number of features ordered by decreasing correlation}
			\label{fig:featureselection}
		\end{figure*}

Table \ref{tab:fcffnnFullDataset} reports high accuracy, confirming the hypothesis of soft error detection based on the collected features. Nevertheless, the accuracy alone may be misleading. The F1 score is lower than the accuracy. This is mainly due to the recall, which drops due to the \ac{HTDR}. To confirm the quality of the model, \autoref{fig:featureselection} investigates the performance of different models trained using an increasing number of features starting from the top correlated ones. In terms of accuracy, using just a small subset of features already enables the detection of 80\% of the faults, and adding more features does not lead to significant improvements.
However, these results consistently show controversial results with the F1 score. By performing a more detailed analysis, we could directly relate this with the complexity of the benchmark and, in particular, its control flow and how the \acs{HTDR} populate the data. As a further confirmation, an attempt to train a model on a reduced dataset focused on the \acs{HTDR} reported deficient performance. This analysis suggests that microarchitecture-level features can only partially detect soft errors, contradicting previous works such as da Rosa et al. \cite{da-Rosa:2019aa}. This problem seems less severe in control-flow intensive benchmarks such as \texttt{qsort}, \texttt{dijkstra}, \texttt{susan} and \texttt{sha} while exploding in simple linear algorithms like \texttt{basicmath} and \texttt{bitcount}. This suggests that data-related features are probably required in addition to the microarchitecture attribute to cover the gap.

After analyzing the cumulative dataset, the analysis moves to new models considering the time dimension using data collected at different checkpoints. 

The first one accounts for a flat dataset composed of $\mathrm{N_{features} \times N_{checkpoints}}$ features and trains the same FC-FFNN model used before. \autoref{tab:fcffnnMulticp} reports no significant gain in performance metrics, thus confirming what was observed in \autoref{subsec:dataAnalysis}, i.e., the temporal dimension is not improving the results but allows for earlier detection.

		\begin{table}[h]
			\centering
		\caption{\ac{FC-FFNN} performance metrics, 10 checkpoints}
		\label{tab:fcffnnMulticp}
			\begin{tabular}{ |c|c|c| } 
				\hline
				Benchmark & Accuracy & F1 score \\ 
				\hline
				qsort 		& 92.60 & 83.07\\
				\hline
				dijkstra 	& 91.67 & 59.00\\
				\hline
				susan 		& 87.95 & 69.67\\
				\hline
				sha 			& 87.14 & 68.24\\
				\hline
				bitcount 	& 90.19 & 58.42\\
				\hline
				basicmath 	& 96.38 & 62.29\\
				\hline
			\end{tabular}		
		\end{table}

A second model, based on \ac{LSTM}, including temporal information by design, was trained on the time-expanded dataset. \autoref{tab:lstmMulticp} reports the performance of the LSTM model, confirming no gain on the different metrics. 

		\begin{table}[h]
		\caption{200 \ac{LSTM} blocks-25FC-2FC performance metrics, 10 checkpoints}	\label{tab:lstmMulticp}
			\centering
			\begin{tabular}{ |c|c|c| } 
				\hline
				Benchmark & Accuracy & F1 score \\ 
				\hline
				qsort 		& 91.62 & 81.09\\
				\hline
				dijkstra 	& 90.25 & 68.98\\
				\hline
				susan 		& 87.83 & 82.50\\
				\hline
				sha 			& 86.85 & 69.08\\
				\hline
				bitcount 	& 90.08 & 54.74\\
				\hline
				basicmath 	& 95.61 & 53.64\\
				\hline
			\end{tabular}
		\end{table}

		\begin{table*}[!htb]
			\centering
		\caption{FC-FFNN, model evaluation at each checkpoint}
		 \label{tab:multipleFcffnn}
			\begin{tabular}{ |c|c|c|c|c|c|c|c|c|c|c|c|c| } 
				\hline
								& \multicolumn{2}{|c|}{qsort}  			&  	\multicolumn{2}{|c|}{djkstra}  			&  	\multicolumn{2}{|c|}{susan}     		&  	\multicolumn{2}{|c|}{sha}			&  	\multicolumn{2}{|c|}{basicmath}		&  \multicolumn{2}{|c|}{bitcount}			\\ 
				\hline
				Temporal step 	& Acc. 		& F1 score 	& Acc. 		& F1 score 	& Acc. 		& F1 score 	& Acc. 	& F1 score 	& Acc. 	& F1 score 	& Acc. 	& F1 score  \\ 
				\hline
1	&80.09	&		39.26	&		81.07	&		13.72	&		72.62	&		9.13	&		73.48	&		- &	83.99	&		- &	92.86	&		- \\
2	&87.78	&		69.99	&		85.96	&		44.79	&		75.06	&		13.36	&		74.72	&		9.55	&		83.68	&		- &	93.67	&		- \\
3	&87.38	&		67.13	&		90.40	&		67.84	&		80.56	&		40.83	&		74.61	&		8.61	&		84.70	&		10.42	&		93.36	&		- \\
4	&90.53	&		78.02	&		90.12	&		60.36	&		87.08	&		68.12	&		75.86	&		17.38	&		85.54	&		21.63	&		94.31	&		24.46	\\
5	&91.44	&		80.15	&		90.70	&		64.65	&		87.86	&		68.93	&		80.79	&		34.31	&		86.93	&		30.38	&		94.78	&		36.54	\\
6	&91.08	&		80,00	&		91.38	&		72.14	&		87.52	&		67.57	&		83.27	&		42.56	&		88.18	&		40.20	&		95.78	&		51.40	\\
7	&91.23	&		80.13	&		90.82	&		70.14	&		87.08	&		66.86	&		87.25	&		65.92	&		88.90	&		45.24	&		95.93	&		58.28	\\
8	&91.72	&		81.56	&		90.44	&		69.99	&		87.84	&		69.86	&		86.77	&		62.85	&		89.73	&		49.47	&		96.33	&		63.91	\\
9	&91.80	&		80.86	&		90.55	&		69.46	&		88.24	&		69.65	&		86.94	&		68.62	&		91.26	&		42.27	&		96.35	&		60.21	\\
10	&91.95	&		81.39	&		91.24	&		73.06	&		88.07	&		69.69	&		86.83	&		67.66	&		89.99	&		57.10	&		96.36	& 61.93\\			
				\hline
			\end{tabular}
		\end{table*}
		
	Since the Fault detection latency can be crucial in safety-critical applications, an FC-FFNN model was then trained on the cumulative dataset and tested on cumulative data available at different checkpoints to understand the early detection of faults. Table \ref{tab:multipleFcffnn} shows the results obtained with ten checkpoints. The F1 score with no values (-) is due to insufficient data to evaluate precision and recall. Considering that the fault injection is uniformly distributed in time, results confirm that error detection could be achieved without waiting for the end of the program execution. This is important to minimize the error detection latency. As expected, the model's performance drops during the early checkpoints and rises while data are collected.

%
%

%% file: Sections/conclusions.tex
\section{Conclusions}
\label{sec:conclusions}
	
This paper performed a data-driven investigation to answer the fundamental question of using AI-powered hardware soft error detectors: are micro-architectural features able to explain faulty executions? The results of the analysis were controversial. While in terms of accuracy, results suggest a positive answer, the presence of \acp{HTDR} suggests that pure microarchitectural attributes are insufficient, especially for simple tasks. Additional features are probably required if the corruption is limited to the data domain. In conclusion, the results in this paper are preliminary. It is clear that further investigation is needed, both considering additional benchmarks and investigating a new set of features.